\documentclass[fleqn,usenatbib]{mnras}

\usepackage{newtxtext,newtxmath}

\usepackage[T1]{fontenc}
\usepackage{ae,aecompl}

\usepackage{graphicx}	
\usepackage{amsmath}	
\usepackage{amssymb}	

\title[Dynamical friction of globular clusters]{The influence of dark matter halo on the stellar stream \\asymmetry via dynamical friction}

\author[Kipper, Tenjes, H\"utsi, Tuvikene, Tempel]{
Rain Kipper$^{1}$\thanks{E-mail: rain.kipper@ut.ee}, 
Peeter Tenjes$^{1}$, 
Gert H\"utsi$^{1,2}$, 
Taavi Tuvikene$^{1}$, 
Elmo Tempel$^{1}$ 
\\
$^{1}$Tartu Observatory, University of Tartu, Observatooriumi 1, 61602 T\~oravere, Estonia\\
$^{2}$National Institute of Chemical Physics and Biophysics, Akadeemia tee 23, 12618 Tallinn, Estonia
}

\date{Accepted 2019 May 9. Received 2019 April 12; in original form 2019 February 12}

\pubyear{2019}

\begin{document}
\label{firstpage}
\pagerange{\pageref{firstpage}--\pageref{lastpage}}
\maketitle

\begin{abstract}
We study the effect of dynamical friction on globular clusters and on the stars evaporated from the globular clusters (stellar streams) moving in a galactic halo. Due to dynamical friction, the position of a globular cluster (GC) as a stream progenitor starts to shift with respect to its original position in the reference frame of initial GC orbit. {Therefore the stars that have evaporated at different times have different mean position with respect to the GC position. } This shifting results in a certain asymmetry in stellar density distribution between the leading and trailing arms of the stream. The degree of the asymmetry depends on the characteristics of the environment in which the GC and the stream stars move. As GCs are located mainly in outer parts of a galaxy, this makes dynamical friction a unique probe to constrain the underlying dark matter spatial density and velocity distributions. For a GC NGC~3201 we {compared our theoretical shift estimates with available observations}. Due to large uncertainties in current observation data, we can only conclude that the derived estimates have the same order of magnitude.
\end{abstract}


\section{Introduction}
Stellar streams that form from the evaporating globular clusters allow to constrain the mass distribution models of the Milky Way (MW) galaxy. For example, by using the general shape of the stream orbits \citet{Bovy:2016} has determined the flatness of the mass distribution of the {inner $20$ kpc of the} MW. Since the stars that evaporate from globular clusters have no preferred direction (in cluster's frame of reference), initially the streams formed at Lagrangian points L1 and L2 should be symmetrical.

The stellar linear density along a stream is not uniform, because epicyclic motion of evaporated stars define fluctuations in density, but these fluctuations can be estimated \citep[see e.g.][]{Capuzzo_Dolcetta:2005, Kupper:2008, Kupper:2012}. In addition, density gaps may occur, as streams may interact with massive molecular clouds when passing through the galactic disc \citep{Dehnen:2004, Amorisco:2016}. An especially interesting topic is the possible interaction of streams with the dark matter (DM) subhaloes and the study of resulting gaps in density distribution of the streams and corresponding perturbations in velocities of stream stars \citep{Ibata:2002, Carlberg:2012, Erkal:2015, Erkal:2016, Erkal:2017, Bovy:2017}. 

According to the $\Lambda$CDM model, at present epoch, the DM subhaloes provide up to $\sim 10\%$ of the overall DM halo mass for a MW size galaxy with a specific (approximate power-law) mass spectrum \citep{Diemand:2008, Springel:2008}. Power spectrum of density distribution fluctuations may make it possible to constrain the mass distribution of DM subhaloes \citep{Bovy:2017}, which in turn may constrain possible warm DM particle masses \citep{Viel:2013, Pullen:2014, Richings:2018} or charged DM particle masses \citep{Kamada:2013}. 

The asymmetry of the mass distribution of the MW (a rotating bar) and corresponding non-stationarity creates short streams \citep{Hattori:2016,Price-Whelan:2016} and an asymmetry between the two branches of the stream \citep{Pearson:2017}. Simulating Palomar $5$ stream asymmetry \citet{Pearson:2017} found that for a certain bar pattern speed the observed asymmetry can be produced. 

In this work we study how the dynamical friction due to stellar and DM halo environment creates  asymmetry in stellar streams. Dynamical friction effect (deceleration) is proportional to the mass of the body and to the density of the environment. Thus deceleration is different for the GCs and for the stars in corresponding stellar streams. Position of a progenitor GC in a stream changes with respect to its initial position. This in turn creates a certain asymmetry in the stellar density distribution between the leading and trailing parts of the stream. The deceleration of the GC can be calculated by measuring this asymmetry. 

When the deceleration is known it is possible to calculate the density of the environment. If the environment consists of different ingredients (gas particles, stars, diffuse DM particles, DM subhaloes), and if the gas and stellar densities are known from independent measurements, it is possible to estimate the DM density and potentially discriminate the smooth DM component from the more dense DM subhaloes. 

Outline of the paper is as follows. In Sect.~\ref{sec:stream_form} we discuss the formation of a stellar stream and present a model to calculate the shift of a GC with respect to its original position within previously evaporated stars due to dynamical friction. In Sect.~\ref{sec:MW} we give the details of the MW model used in subsequent calculations. Sect.~\ref{sec:results} presents our main results. Results are given for a large sample of GCs and thereafter we compare them with available stream data. Sect.~\ref{sec:discussion} is left for discussion and conclusions.

\section{Stellar streams of globular clusters}\label{sec:stream_form}

\subsection{Formation of a stream}

In an axisymmetric galaxy, all the test particles (stars, GCs etc.) move in orbits that conserve integrals of motion, such as the energy and the $z$-directional angular momentum $L_z$. Interactions between the galactic objects can change their integrals of motion, which also changes their orbits {(see e.g. \citet{Sellwood:2013})}. 

In case of the dynamical evolution of a GC interactions between the stars within the GC cause their evaporation from the GC and subsequent formation of stellar streams { \citep{Fall:1977,Gnedin:1997,Odenkirchen:2001}.} At the moment stars evaporate from the GC (primarily through the Lagrangian points $L_1$ and $L_2${, see \citet{Kupper:2008,Kupper:2012} and \citet{Bowden:2015}}) they have a shift in positions and velocities with respect to the centre-of-mass values of the GC. Thus, the integrals of motion for the GC and corresponding evaporated stars are slightly different, causing GC and its stream stars to move along slightly different orbits { \citep{Eyre:2011,Bovy:2014,Carlberg:2017}}. Small differences in velocities between individual stream stars at Lagrangian points result in slightly different orbits for them, hence, the stream has also some thickness. 

The evaporation through two Lagrangian points results in two stream arms -- the leading and the trailing arm. As GCs can be assumed to be symmetrical, with their dimensions very small in comparison to the parent galaxy, the gravitational potential of the galaxy close to the GC can be approximated with a linear function. In this case, both Lagrangian points are at the same distance from the centre of the GC and intrinsically the stellar stream is symmetric -- leading and trailing arms have statistically the same distribution of stars with respect to the GC centre. {\citet{Erkal:2016b} showed that precession and nutation that cause non-uniformity of the stream come forth for longer streams. }

In a non-stationary galaxy stream arms are not symmetrical. Non-stationarity may be caused by large-scale interactions or by a rotating bar {(e.g. \citet{Pearson:2017})}. However, small non-stationarities may also be caused by, e.g., dynamical friction effects, leading to slight incremental changes in GC orbits over time.

In this paper we analyze the asymmetry of stellar streams caused by small changes of GC orbit (or corresponding integrals of motion) between different evaporation events. Changes in a GC orbit are primarily caused by the dynamical friction in a host galaxy. 

\begin{figure}
    \includegraphics[width=\columnwidth]{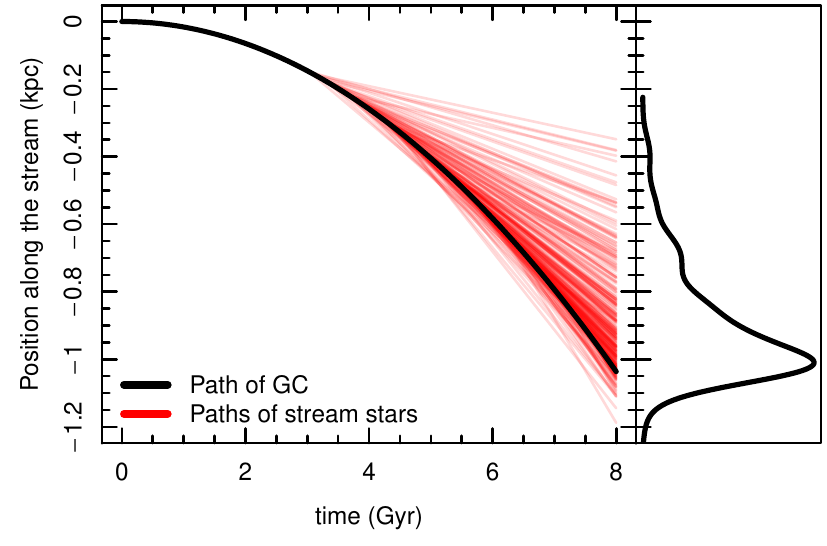}
    \caption{{An illustration about t}he shift of GC position due to dynamical friction along a stream with respect to its initial position as a function of time. In the left hand panel the black solid line depicts GC position shift, thin red lines denote individual evaporated stars. It is seen that at every moment the projected positions of stream stars are not symmetrical with respect to the GC position. Asymmetry increases in time. In the right hand panel the resulting number density distribution of evaporated stars along the stream at $t=8$~Gyr is given. Density distribution asymmetry with respect to the GC position at $t=8$~Gyr is clearly visible.}
    \label{fig:illustration}
\end{figure}

In case of a stationary galaxy and a symmetric stream, the position of the progenitor GC and the overall centre of the symmetry of stream stars are the same and remain the same in time. However, if the GC orbit changes between ejections of stars compared to the initial orbit, subsequent stream stars will be ejected symmetrically with respect to the new orbital position of the progenitor GC. As a result, the mean position of all stream stars does not coincide now with the GC. The shift between the centre of the GC and the centre of stream stars changes in time. Evolution of the shift for a simple model including the dynamical friction effect {(a constant deceleration)} is illustrated in Fig.~\ref{fig:illustration}. {Deceleration value for this illustration was not calculated from friction model but was taken simply to be $0.03$~km/s/Gyr.} Velocities of evaporated stars were taken from a Gaussian distribution with the same mean value and dispersion as the GCs orbital velocity and a typical GC velocity dispersion. Since dynamical friction affects only GC orbit and keeps the orbits of the evaporated stars intact\footnote{The dynamical friction on stream stars is completely negligible, $4-5$ orders of magnitude smaller.}, time-evolving shift between the mean position of evaporated stars and GC location appears (left-hand panel in Fig.~\ref{fig:illustration}). As a result, the averaged number densities of stream stars are  not symmetrical between the two arms of the stream (right-hand pane in Fig.~\ref{fig:illustration}).

Dynamical friction has small effect on a GC orbit. For this reason, we characterize the changes of a GC orbit and positions of objects along the orbit simply as shifts projected to the initial GC orbit -- shifts along the orbit. These shifts along and with respect to the initial GC orbit are denoted as $\Delta x$.

\subsection{Dynamical friction}

Dynamical friction is a drag force on a massive object moving through the field of lighter objects and collecting a temporary mass wake behind it \citep{Chandrasekhar:1943}. The wake applies force to the massive object and decelerates it. 

A general formula for the force per unit mass caused by the dynamical friction is given as \citep[see e.g.][]{Binney:2008}
\begin{equation}
    F =  -4\pi G^2 M \ln{\Lambda}\rho({\bf x})\iiint f({\bf x,v'})\frac{{\bf v'} - {\bf v_{\rm GC}}}{|{\bf v'} - {\bf v_{\rm GC}}|^3}{\rm d}{\bf v'}, \label{eq:fric_general}
\end{equation}
where ${\bf v_{\rm GC}}$ and $M$ are the velocity and the mass of the massive object (e.g. GC), respectively, $\rho$ is the spatial density of field objects and $f$ is their phase space density normalized to one, $\ln\Lambda$ is the Coulomb logarithm and $G$ is the gravitational constant. The matter contributing to the dynamical friction consists mainly of stars and the DM halo particles. In case of the isotropic velocity distribution of field objects, this equation simplifies considerably, and is called Chandrasekhar's formula \citep[see][]{Chandrasekhar:1943}. 

From Eq.~\eqref{eq:fric_general} we see that the effect of dynamical friction is stronger for more massive GC and for the ones that move with similar velocities as the field objects (denominator under the integral in Eq.~\eqref{eq:fric_general} is small).

\subsection{Combined measurable quantities}\label{sec:donkey}
{The orbital segments of stream stars and of the parent GC are approximately parallel to each other. For simplicity, we measure distances only along the GC orbit and positions of stars along the stream orbit are projected to the GC orbit.} Let $\Delta x$ be a shift of a GC and evaporated stars with respect to the initial GC position projected to the initial GC orbit. Hence, the shift $\Delta x$ results mainly from velocity differences between different orbits and in calculations we approximate the shift $\Delta x$ based on velocities along the GC initial orbit.

Thus, the shift $\Delta x$ at time $t$ can be calculated as
\begin{eqnarray}
    \Delta x(t) &=& \int\limits_{t_{\rm initial}}^{t} \Delta v(t') {\rm d}t' , \label{eq:dx_leidmise_valem}\\
    \Delta v(t) &=& \int \limits_{t_{\rm initial}}^{t} {a}(t') {\rm d}t',\label{eq:dv_leidmise_valem}
\end{eqnarray}
where $\Delta v(t)$ is the velocity difference at time $t$ and $a$ is the acceleration. Quantities $\Delta x$, $\Delta v$, and $a$ are measured along the orbit of GC and can be calculated from the total force per unit mass $\bf F$. {The forces were calculated based on orbit after its integration, hence the friction was not included in orbit integration. This is justified since the first-hand estimation of dynamical friction does not change the orbit noticeably.}

To estimate the acceleration $a$ from $\bf F$ we have to take into account the response of velocity to the torque via changes of angular momentum $L_z$. {The origin of the torque is the dynamical friction.}  This can be approximated by the so-called `donkey effect' (see \citealt{Binney:2008})
\begin{eqnarray}
    {\bf a}_{\varphi,\theta} &=& \frac{A}{B}{\bf F}_{\varphi, \theta},\\
    {\bf a}_{r} &=& {\bf F}_{r},
\end{eqnarray}
where $\varphi,\theta,r$ are spherical coordinates and $r$ measures the distance from the centre of host galaxy. In the last equation we have assumed the usual Oort constants and circular velocity ($v_c$) denotations
\begin{eqnarray}
    A(R) &=& \frac12\left( \frac{v_c}{R} - \frac{{\rm d}v_c}{{\rm d}R} \right) , \\
    B(R) &=& -\frac12\left( \frac{v_c}{R} + \frac{{\rm d}v_c}{{\rm d}R} \right) , \\
    v_c &=&  \sqrt{R\frac{{\rm d}\Phi}{{\rm d}r}}.
\end{eqnarray}
Here $R$ is the distance from the centre of a galaxy in the Galactic plane and $\Phi$ denotes gravitational potential. In case of flat rotation curve $A/B$ reduces to $-1$, in case of adopted Galactic potential (see Sect.~\ref{sec:MW}), $A/B$ varies between $-0.2$\dots$-1.5$.

 The acceleration $a$ used in Eq.~\eqref{eq:dx_leidmise_valem} is the projection of $\bf a$ to the GC orbit at a given point.

\section{Description of the Milky Way model}\label{sec:MW}

To calculate the asymmetry in a stellar stream density distribution, described in the previous section, we need to know the mass density distribution $\rho$ and the phase space density distribution (in fact, the velocity distribution) $f$ for a region of the Galaxy where the stream is located. 

\subsection{Spatial densities and velocity distributions of Galactic components}

To calculate the orbits of progenitor GCs we used {\tt galpy} python package developed by \citet{Bovy:2015} with {\tt MWPotential2014}. To calculate the dynamical friction effect from Eq.~(\ref{eq:fric_general}) we need density and velocity distributions of corresponding individual components. Although the components in {\tt MWPotential2014} included a bulge, a Miyamoto-Nagai disc and a NFW dark halo, the first two components do not contribute in outer regions of the MW, where the dynamical friction effects are calculated. However, a stellar halo may contribute to the dynamical friction. Thus, we added a stellar halo from a model by \citet{Juric:2008}. 

We also need to know the velocity distribution function at the orbital points of GCs. Dynamical friction results from the stars of stellar halo and from the DM particles. For the stellar halo all three velocity dispersion components were taken from \citet{Bird:2018}. Unfortunately, this kind of data is not available for the DM halo. To derive velocity dispersions for the DM halo we solved Jeans equations. Two forms for the corresponding velocity ellipsoids were assumed: (i) an isotropic velocity distribution, and (ii) a distribution with constant anisotropy. The solutions for the Jeans equations for these two cases are
\begin{equation}
    \sigma_r^2(r) = \frac{1}{\rho_{\rm DM}r^\gamma}\int\limits_r^\infty \rho\frac{{\rm d}\Phi}{{\rm d} x}x^\gamma{\rm d}x,\label{eq:Jeans_solution}
\end{equation}
for constant anisotropy case, where $\gamma=2-2\sigma^2_{\theta,\phi}/\sigma^2_r$ is an anisotropy parameter, and 
\begin{equation}
    \sigma^2(r)=\frac1{\rho_{\rm DM}(r)} \int\limits_r^{\infty}\rho_{\rm DM}(x) \frac{{\rm d}\Phi}{{\rm d}x} {\rm d}x
\end{equation}
for the isotropic case. The anisotropy parameter in Eq.~(\ref{eq:Jeans_solution}) was assumed to be the same as the mean anisotropy of the stellar halo, $\sigma_{\theta,\phi}/\sigma_r = 0.62$, taken as the average from the halo profile of~\citet{Bird:2018}. 

\begin{figure}
    \centering
    \includegraphics[width=\columnwidth]{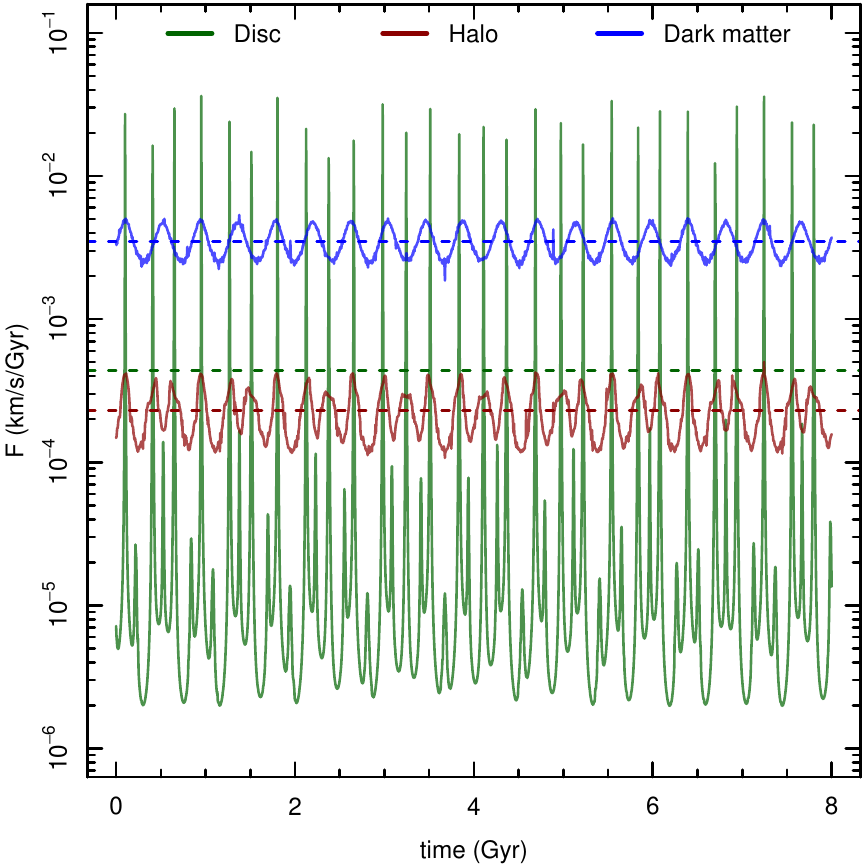}
    \caption{{Contribution of different galaxy components to the overall dynamical friction for a GC NGC 5024. The mean contribution from stellar components (green and red lines) are about an order of magnitude smaller than from the dark matter component (blue line). Horizontal dashed lines show average contribution of the corresponding component. }}
    \label{fig:fric_components}
\end{figure}

A typical GC orbit is mostly located in the outer regions of the galaxy, where the DM halo dominates the density distribution. The contributions of stellar disc and stellar halo to the dynamical friction are rather small. We tested this by selecting their average velocity distribution parameters according to \citet{Hawthorn:2016}. The result is presented in Fig.~\ref{fig:fric_components} for a typical GC with mean distance of  $r = 35$~kpc from the Galactic centre. {The stellar halo and disc provide about an order of magnitude smaller contribution integrated over time than the DM}. Hence, in subsequent dynamical friction calculations we ignore the contributions of both stellar components. 

\subsection{Globular clusters in the Milky Way}

GCs analysed in this paper and their parameters were taken from a compilation by \citet{Baumgardt:2018}.\footnote{Data available from web page \url{https://people.smp.uq.edu.au/HolgerBaumgardt/globular/}} For each GC we calculated their orbits and shift $\Delta x$.   

The Coulomb logarithm in Eq.~(\ref{eq:fric_general}) determines the region along the GC path that is contributing to the dynamical friction. We used the value $\ln\Lambda = 5.8$ by assuming a maximum impact parameter of $1\,{\rm kpc}$, and a minimum impact parameter $3\,{\rm pc}$ corresponding to the typical half-light radius of GCs.

\section{Results}\label{sec:results}

We calculated the effect of dynamical friction for the GCs in \citet{Baumgardt:2018} selection. In Section~\ref{sec:example} we describe the effect of dynamical friction in detail by using one GC as an example. In Section~\ref{sec:measurements} we present the results for all GCs and in Section~\ref{sec:gcs} we discuss {some} GCs in more detail.

\subsection{Dynamical friction effect for a typical GC}\label{sec:example}
\begin{figure*}
    \centering
    \includegraphics[width=\textwidth]{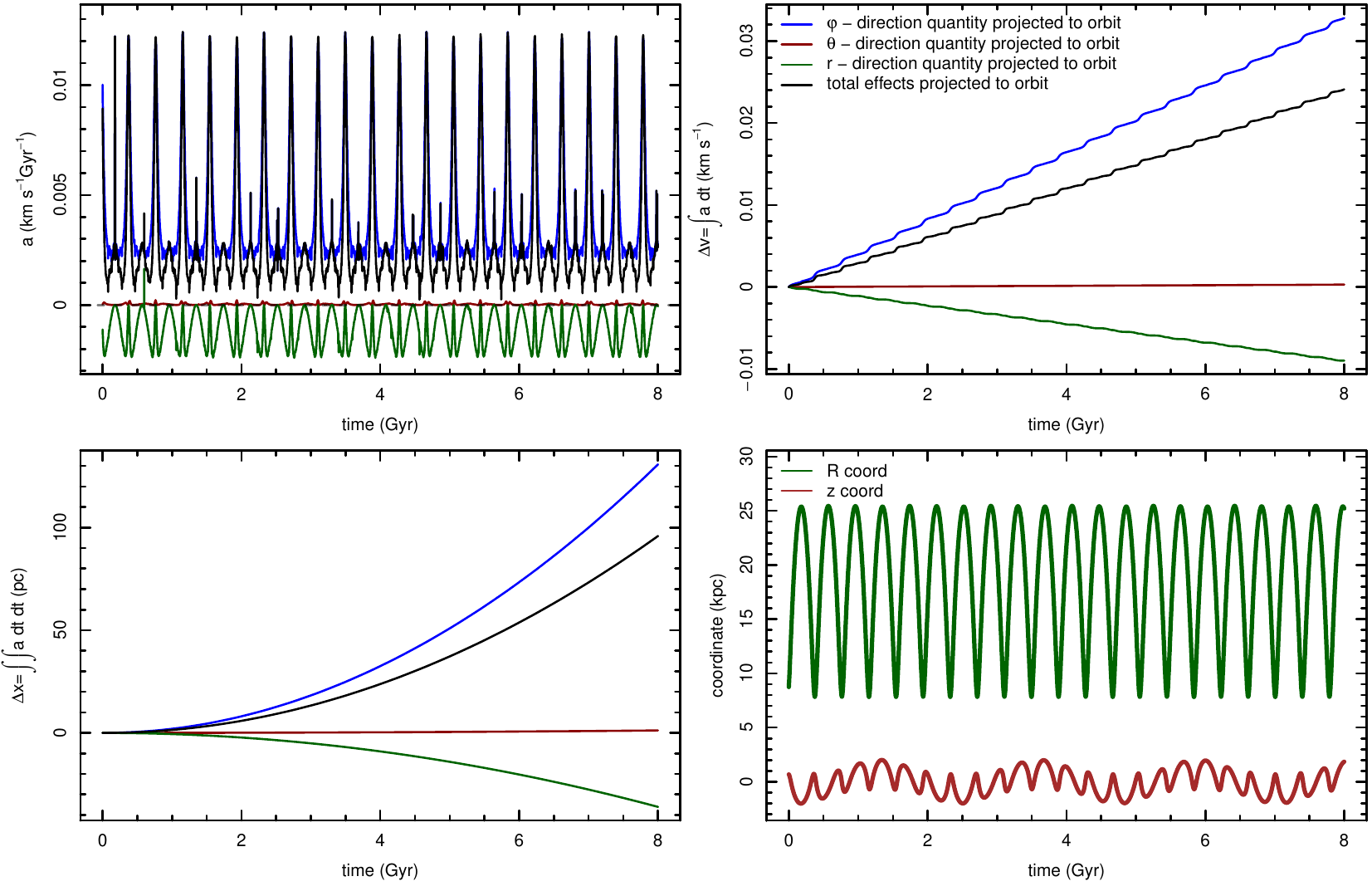}
    \caption{Evolution of calculated acceleration values (top left), velocity shift components (top right, see Eq.~\eqref{eq:dv_leidmise_valem}) and position shift along the orbit (bottom left, see Eq.~\eqref{eq:dx_leidmise_valem}) for a sample GC NGC 3201. Colour coding for these three panels is the same. Bottom right panel shows the shape of the orbit (radius and height from the Galactic plane). Calculations were made assuming an isotropic velocity distribution. GC orbit is calculated over 8~Gyr assuming that MW potential does not change during this time. }
    \label{fig:ngc3201}
\end{figure*}

We selected a well observed GC NGC~3201 as an example for which we present possible stellar stream asymmetry sources in detail. Fig.~\ref{fig:ngc3201} shows the calculated decelerations, resulting velocity and position shifts along the original GC orbit over the last 8~Gyr. Both the total deceleration (black lines) and the contribution from individual deceleration components (green for radial component, red and blue for tangential ones) are given. Periodicity seen in deceleration values corresponds to different locations of the GC in the Galaxy. During the pericentre passages, the deceleration has prominent peaks due to increased DM densities. During the apocentre passages, there are secondary peaks caused by the slower motion of the GC (or even nearly standing still along the radial direction). This effect results from the integral part of Eq.~\eqref{eq:fric_general}. It means also that the radial acceleration contribution to the dynamical friction (projected to the GC orbit) is zero during the apocentric passages. 

It is also seen that acceleration directions (positive or negative) and shift values along the orbit are different for different acceleration vector components. This is caused by the donkey effect (see Section~\ref{sec:donkey}). Depending on the nature of GC orbit (radial or circular orbits), the shifts can be positive or negative with respect to the orbital motion {since the orbit determines how strongly the donkey effect affects the shift}. Hence, it depends on the nature of a GC orbit whether the leading or the trailing arm of a stellar stream is longer. For NGC~3201, the trailing arm is longer than the leading arm. Fig.~\ref{fig:ngc3201} shows that $\theta$-component (red line) accelerations are very small and thus have nearly negligible contribution to the coordinate and velocity shifts. This is not surprising since NGC~3201 orbit is quite close to the Galactic plane and oscillates slowly in vertical direction (see the bottom right panel on Fig.~\ref{fig:ngc3201}). 

An additional point of interest is that the cumulative effect of the dynamical friction (shift $\Delta x$ along the orbit) depends quite strongly on time. Hence, the asymmetry between leading and trailing arms is strongest for stars that are evaporated during earlier times in cluster's dynamical evolution. Due to mass segregation, stars evaporated earlier tend to be less massive and are, unfortunately, relatively {faint} and are difficult to observe. 

\subsection{Dynamical friction effect for the total sample of GCs}\label{sec:measurements}
\begin{figure*}
    \centering
    \includegraphics[width=\textwidth]{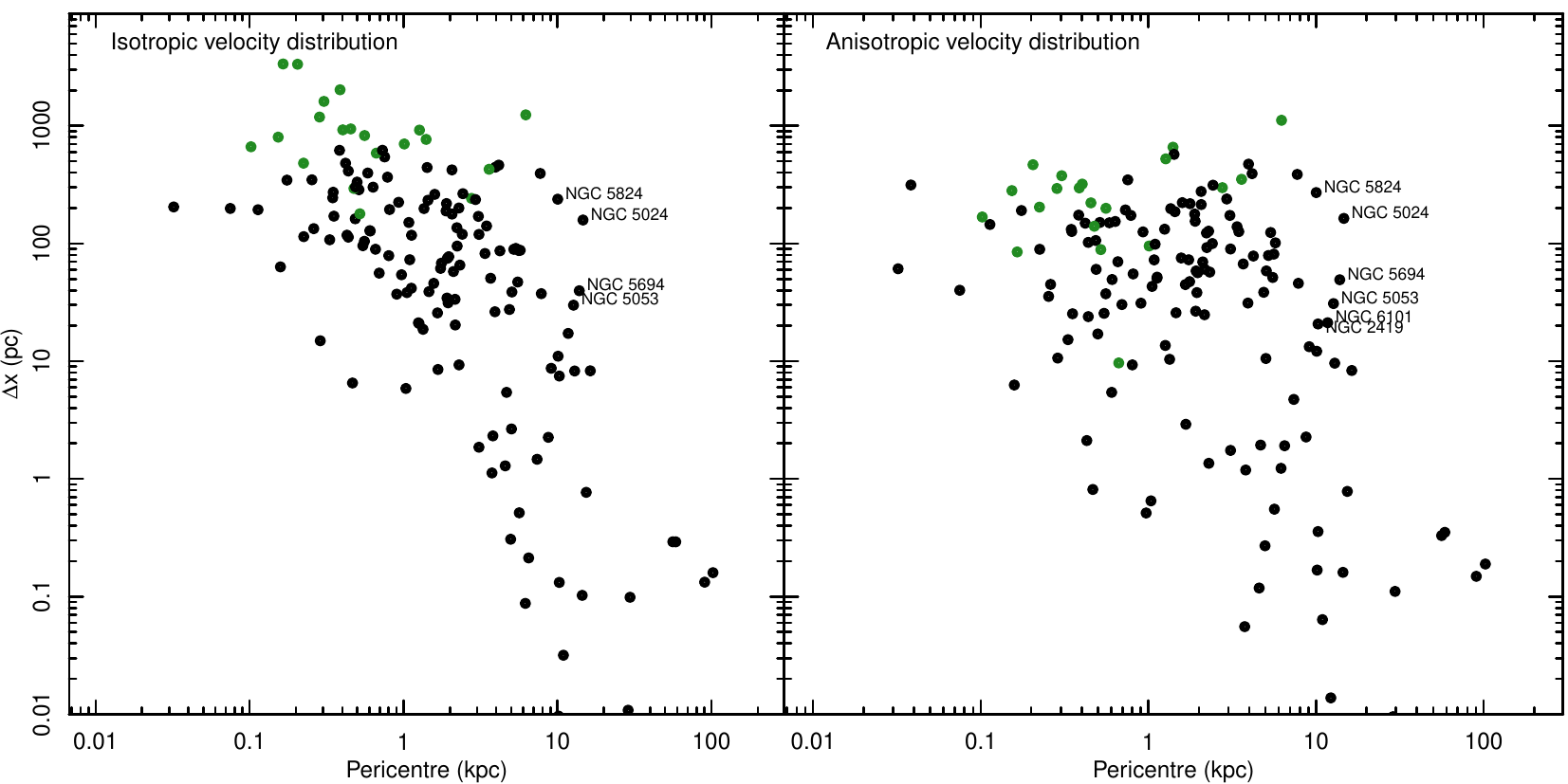}
    \caption{Calculated dynamical friction shift along the orbit for all GCs with known phase space coordinates. To characterize the strength of influence of the Galactic bar on the stream asymmetry, the $x$-axis shows the pericentric distance to the GC orbit. {The calculations assume isotropic velocity distribution for left panel, and anisotropic one for right panel. Some of the most interesting GCs are indicated with labels.} 
    }
    \label{fig:dx5}
\end{figure*}

The effect of dynamical friction (shifts along the orbit) {as a function of GC pericentre distance} is presented in Fig.~\ref{fig:dx5}. The shifts are calculated for a $5\,$Gyr time span {for all GCs from Baumgardt catalogue \citep{Baumgardt:2018} with known phase space coordinates. The Figure shows that calculations assuming isotropic or anisotropic velocity distribution give similar results. Shifts calculated by assuming a constant anisotropy fixed to stellar halo value are smaller when compared to the isotropic case.} GCs with small pericentric distances might be influenced by the MW bar. Hence, at present (when the bar effects cannot be modelled reliably) the shift estimates for these GCs are not reliable.

{Labelled} GCs at the upper right part of Fig.~\ref{fig:dx5} are the best candidates to detect the dynamical friction effect from dark matter halo. These GCs are in orbits with large pericentre values and the effect of dynamical friction for these GCs is the largest. In the next subsection we will discuss some of these GCs in more detail. {Large shift is not the only criterion for observational detection of dynamical friction effects, because GC location at a low Galactic latitude or at a large distance from us makes observing streams challenging.}

{Another subset of GCs in Fig.~\ref{fig:dx5} is marked with green symbols and they denote the GCs that have shifts which translate to largest angular distance in the sky. These GCs are mostly located close to the Sun. Although from observational aspect their stream asymmetries are easier to detect than ones having large shift, their position is too close to the bar and possible asymmetry detection does not allow us to conclude that the dominant cause of the shifts is dynamical friction.}

{The $5\,$Gyr shifts along the orbit for all analysed GCs are listed in Table~\ref{tab:app1}. There we also present angular shifts, i.e. shifts projected from orbital arc to the sky plane. The angular distance estimate includes the projection of orbital arc perpendicular to line of sight, but does not include orbital velocity inhomogenities.}

\subsection{Specific GCs}\label{sec:gcs}

{In this Section we discuss GCs that are most promising for pure dynamical friction analysis. We included only GCs for which we found relevant references in literature.}

\begin{figure*}
    \centering
    \includegraphics[width=\textwidth]{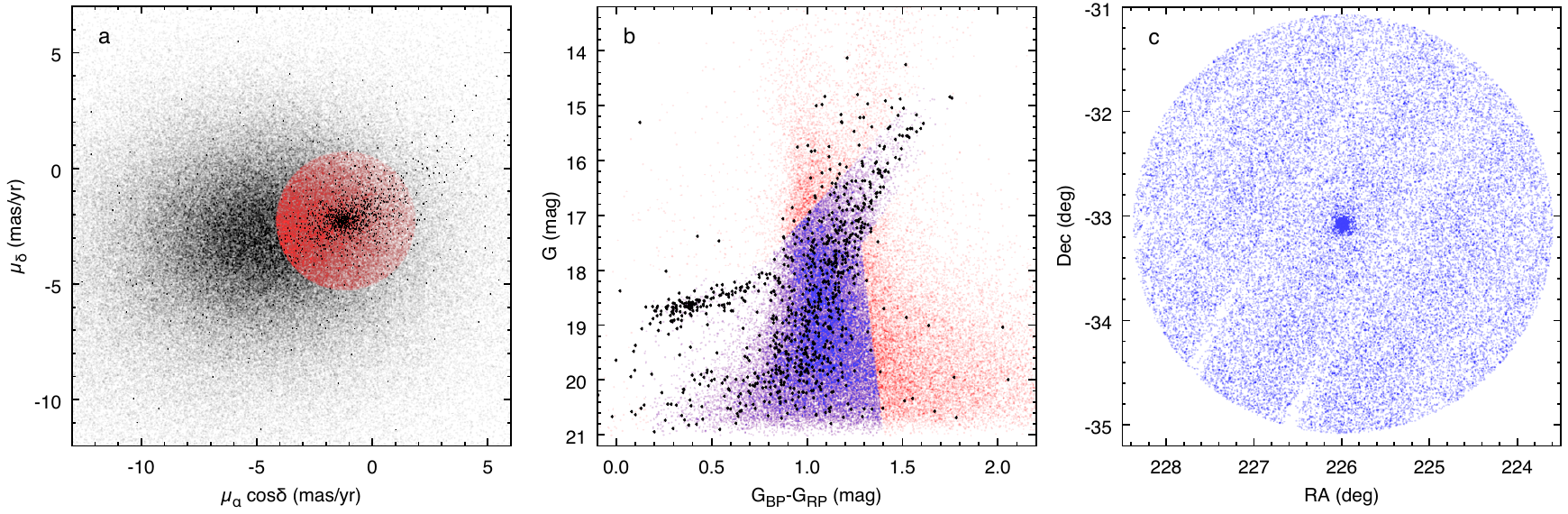}
    \caption{{Search for a stream within two-degree radius from the globular cluster NGC 5824. Panel a shows proper motions of all stars in the area (grey dots). Selected stars with proper motions similar to the GC are plotted with red dots and overlaid with GC stars (black dots), taken from an area with 3-arcmin radius from the cluster centre. The colour-magnitude diagram is shown in panel b: red dots denote all stars from the proper-motion cut, blue dots denote selected stars. The selection was based on the location of GC stars, overlaid with black dots. The sky plot of the final selection is shown in panel c.}}
    \label{fig:ngc5824_multi}
\end{figure*}

\subsubsection{NGC 3201}
Based on $13$ RR Lyra stars {compiled by \citet{Kundu:2018} there is a} shift between the GC and GC stream centre to be $18\pm23\,{\rm pc}$ {(this error includes only statistical one, and does not include any of the selectional sources)}. Our dynamical friction model estimate (see Fig.~\ref{fig:ngc3201}) is $25\dots43 \,{\rm pc}$ (for $5$ Gyr), which is in good accordance with the measured value. {In principle, the observed value is within the errors consistent also with no shift.} 

\subsubsection{Palomar 5}
GC Palomar 5 has long and prominent leading and trailing streams. Palomar 5 might be in resonance with the MW bar \citep{Pearson:2017}, thus unfortunately not eligible for the friction-only analysis. In addition, Palomar 5 has too small mass for a noticeable dynamical friction effect{ from dark matter halo}. 

\subsubsection{NGC 5824, NGC 5024, NGC 5694, NGC 2419, NGC 3201, NGC 6101}
{GCs NGC~5824, NGC~5024, NGC~5694, NGC~2419, NGC~201 and NGC~6101 have strong potential to exhibit detectable effects from pure dark matter dynamical friction.}
We searched for stellar streams in the neighbourhoods of {these GCs}, using data from the Gaia DR2 catalogue \citep{Gaia:2016, Gaia:2018}. {The search process in the case of NGC 5824 is illustrated in Fig.~\ref{fig:ngc5824_multi}.} First, we selected stars that had proper motions similar to the respective GC. Next, we plotted the colour-magnitude ($G_\mathrm{BP}-G_\mathrm{RP}$ vs $G$) diagram (CMD) for each GC, and narrowed the star selection in the GC neighbourhood with stars that followed the GC isochrone on CMD. The resulting selections were visually inspected. No streams were apparent around the above-mentioned GCs.

\section{Discussion and conclusions} \label{sec:discussion}

\subsection{Observational aspects}
We estimated how much the effect of dynamical friction influences the asymmetry of GC stellar streams. The constant lagging of a GC behind on its orbit when compared with the stream stars produces a shift of the GC with respect to the previously evaporated stars and thus an asymmetry between the leading and trailing parts of the stream. Thus, the distribution of recently evaporated stars is preferably more lopsided compared to the earlier evaporated stars. From the observational perspective, it is seen as an asymmetry in surface density distributions between the two branches of the stellar stream. 

We calculated shifts ($\Delta x$) of the progenitor GC and its evaporated stream stars with respect to the initial position of the progenitor GC or the centre of the symmetry of the oldest parts of the stream. In most cases, either calculated shifts are too small to be detected within current observational data, or pericentres of GC orbits are too near the MW central bar, which can be another source of asymmetry in observed stellar streams. Although there are numerous hindrances, it should still be possible to study the dark matter by taking these effects into account and provided deep photometric observations of the stream. {The diversity of the distances, evaporation histories and densities of the streams do not allow us to deduce a certain brightness limit for needed observations. }

On the other hand, calculated shifts, depending on the GC orbit and evaporation process details, may be quite large compared to the width of the stream. In case of short stellar streams (recently evaporated stars) the shift $\Delta x$ is more easily measurable. In principle, there exists a possibility that for some GC there might be even only one stream arm, if the radial derivative of the angular velocity in the orbit is small{ and dynamical friction is large}. In a similar case, when the GC has moved backwards and inward with respect to one arm of the stream and still evaporates stars, it might be possible to have two parallel streams. As an example, a {short} double stream structure near GD-1 has been found by \citep{Malhan:2018} (not the longer one found by \citet{GrillmairDionatos:2006}), but another possible explanation of this double structure is simply { a massive perturber \citep{PrinceWhelan:2018, Bonaca:2018}}. 

Dynamical friction and resulting inward migration of a GC orbit is an aspect which should be taken into account also when streams are used to constrain overall properties of the gravitational potential of a parent galaxy. This is especially important when a GC and corresponding stream is moving in an elongated high energy orbit {which allows to probe the friction it had near the apocentre i.e. further away than the current position}. 

{In the present paper we assumed a simple virialized halo. A first order estimate how much results can vary when loosening this assumption can be seen by comparing panels of Fig.~\ref{fig:dx5}. Incorporating of more complicated haloes to our model (e.g. by using the halo from \citet{Helmi:2018}) can be done by including halo substructures to the phase space function $f$ and include their orbital evolution. }

\subsection{DM implications}
Assuming that with future detailed measurements of GCs it is feasible to determine the strength of dynamical friction in several locations over the Galactic halo, one would certainly hope to learn valuable information about the properties of DM -- the dominant form of material filling the halo.

On the one hand, the assumption used for deriving Eq.~\eqref{eq:fric_general} is that an elementary unit of DM is much lighter than the mass of the target object moving in that medium, and thus, it does not important if the medium consists of particle DM (e.g. axions, WIMPs, etc) or even of primordial black holes with masses up to and exceeding the Solar mass-scale -- the instantaneous dynamical friction is determined simply by the density of the medium. On the other hand, the realistic density and velocity-space structure of the DM halo -- DM streams, subhaloes, etc -- is far from uniform and depends crucially on (i) halo assembly history (i.e. on particular small-scale realization of initial fluctuations) and (ii) on DM microphysics (e.g. warm DM (WDM), fuzzy DM, self-interacting DM -- popular alternatives to the standard cold DM (CDM)). For example, in the standard CDM scenario one expects ${\cal O}(10)\%$ of the DM halo to be in the form of substructures~\citep[e.g.][]{2004MNRAS.355..819G}, while for the cases of WDM and fuzzy DM the amount of small-scale substructures is significantly reduced~\citep[e.g.][]{2012MNRAS.424..684S,2017MNRAS.465..941D}.

In order to have a measurable orbital lag of the GC with respect to the ejected stars one has to use long enough sections of the tidal tails, and thus effectively probe the density of the DM medium averaged over the corresponding section of the cluster's orbit. In cases with large amount of substructures the chances of passing through denser DM clumps is enhanced, leading to a boost in dynamical friction, and thus, in principle giving us sensitivity to differentiate between alternative DM models. Of course, this can be done only statistically once one has several detailed GC measurements available over the MW's DM halo.

The power to discriminate between various DM models is further complicated by the fact that a substantial fraction of GCs in Galactic halo, in particular the metal-poor blue population, is not expected to be formed \emph{in situ}, rather being accreted as member systems of the MW's satellite galaxies~{\citep[e.g.][]{Searle:1978, 2017MNRAS.465.3622R}}. Thus, one would expect the orbits of these GCs for quite some time to be correlated with the tidal DM debris left over from the disruption of their parent haloes, this way enhancing the level of dynamical friction. Also, more recently accreted GCs are expected to have significant velocities with respect to the DM halo, such that the effective velocity distribution function entering in dynamical friction calculations can be significantly anisotropic in GC's frame of reference. 

The main goal of this paper was to estimate the size of the GC displacement due to dynamical friction and to further provide initial investigation under what conditions can the effect be realistically measurable. A more detailed investigation regarding the possibility to distinguish between the alternative DM models is beyond the scope of this paper and is left for the future study.

\section*{Acknowledgements}
{We thank the referee for useful suggestions and comments. }
This work was supported by institutional research funding \mbox{IUT26-2} and \mbox{IUT40-2} of the Estonian Ministry of Education and Research. We acknowledge the support by the Centre of Excellence ``Dark side of the Universe'' (TK133) and by the grant MOBTP86 financed by the European Union through the European Regional Development Fund.
This work has made use of data from the European Space Agency (ESA) mission
{\it Gaia} (\url{https://www.cosmos.esa.int/gaia}), processed by the {\it Gaia}
Data Processing and Analysis Consortium (DPAC,
\url{https://www.cosmos.esa.int/web/gaia/dpac/consortium}). Funding for the DPAC
has been provided by national institutions, in particular the institutions
participating in the {\it Gaia} Multilateral Agreement.

\bibliographystyle{mnras}
\bibliography{gc_fric}

\appendix
\section{Complete table of dynamical friction estimates}
\begin{table*}
    \centering
    \caption{{The estimation of shift along the orbit ($\Delta x$) produced by dynamical friction from only dark matter over the 5\,Gyr time span. Angular shift ($\Delta \phi$) shows the angular distance between the GC and stream centre along the orbit as viewed from the Sun. $D_\odot$ denotes the distance of the GC from the Sun. $l$ and $b$ denote the Galactic longitude and latitude of the GC. The calculations are made by assuming either isotropic or constant anisotropic dark matter particle velocity dispersion tensor with the anisotropy value adopted from stellar halo. }}
    \label{tab:app1}
    \begin{tabular}{l|rrrrrrrr}
    \hline
GC & $l$ & $b$ &  $D_\odot$  & $\Delta x$ (isotropic) & $\Delta \phi$ (isotropic) & $\Delta x$ & $\Delta \phi$ & Pericentre \\
 & deg & deg & kpc & pc & deg & pc & deg & kpc \\
 \hline
NGC 6522 &  1.02 & -3.92 & 8.0 & 3352 & 22.73 & 85 & 0.61 & 0.2 \\
NGC 6440 &  7.73 &  3.80 & 8.2 & 3331 & 21.47 & 467 & 3.16 & 0.2 \\
Ter 1 & 357.56 &  0.99 & 6.7 & 2023 & 13.52 & 297 & 2.04 & 0.4 \\
NGC 104 & 305.89 & -44.89 & 4.4 & 1238 & 8.53 & 1114 & 7.71 & 6.2 \\
NGC 6380 & 350.18 & -3.42 & 9.8 & 1611 & 8.31 & 376 & 1.96 & 0.3 \\
NGC 6626 &  7.80 & -5.58 & 5.4 & 827 & 7.51 & 199 & 1.82 & 0.6 \\
NGC 5139 & 309.10 & 14.97 & 5.2 & 766 & 7.48 & 658 & 6.44 & 1.4 \\
NGC 6273 & 356.87 &  9.38 & 8.3 & 1186 & 7.22 & 293 & 1.80 & 0.3 \\
NGC 6401 &  3.45 &  3.98 & 7.7 & 941 & 6.72 & 221 & 1.59 & 0.5 \\
NGC 6266 & 353.57 &  7.32 & 6.4 & 919 & 6.48 & 523 & 3.71 & 1.3 \\
NGC 6544 &  5.84 & -2.20 & 2.6 & 293 & 6.43 & 140 & 3.08 & 0.5 \\
Pal 6 &  2.09 &  1.78 & 5.8 & 801 & 4.68 & 281 & 1.65 & 0.2 \\
NGC 6752 & 336.49 & -25.63 & 4.2 & 427 & 4.63 & 351 & 3.81 & 3.6 \\
NGC 6093 & 352.67 & 19.46 & 8.9 & 663 & 4.25 & 168 & 1.08 & 0.1 \\
NGC 6656 &  9.89 & -7.55 & 3.2 & 242 & 4.04 & 298 & 4.98 & 2.8 \\
NGC 6712 & 25.35 & -4.32 & 7.0 & 481 & 3.96 & 204 & 1.68 & 0.2 \\
NGC 6453 & 355.72 & -3.87 & 11.6 & 923 & 3.89 & 320 & 1.35 & 0.4 \\
NGC 6121 & 350.97 & 15.97 & 2.0 & 179 & 3.56 & 89 & 1.77 & 0.5 \\
NGC 6441 & 353.53 & -5.01 & 11.8 & 700 & 3.39 & 95 & 0.46 & 1.0 \\
NGC 6402 & 21.32 & 14.80 & 9.3 & 586 & 2.64 & 10 & 0.04 & 0.7 \\
NGC 4833 & 303.60 & -8.02 & 6.2 & 286 & 2.62 & 151 & 1.38 & 0.5 \\
NGC 5986 & 337.02 & 13.27 & 10.6 & 481 & 2.54 & 149 & 0.79 & 0.4 \\
NGC 6388 & 345.56 & -6.74 & 10.7 & 442 & 2.25 & 571 & 2.90 & 1.4 \\
NGC 6569 &  0.48 & -6.68 & 10.6 & 422 & 2.23 & 277 & 1.46 & 2.1 \\
NGC 6638 &  7.90 & -7.15 & 10.3 & 619 & 2.04 & 174 & 0.57 & 0.4 \\
NGC 6205 & 59.01 & 40.91 & 6.8 & 397 & 2.03 & 150 & 0.77 & 0.6 \\
NGC 6517 & 19.23 &  6.76 & 10.6 & 619 & 2.02 & 193 & 0.63 & 0.7 \\
NGC 5272 & 42.22 & 78.71 & 9.6 & 393 & 2.01 & 386 & 1.97 & 7.7 \\
NGC 6355 & 359.58 &  5.43 & 8.7 & 413 & 1.99 & 102 & 0.49 & 0.4 \\
NGC 6304 & 355.83 &  5.38 & 5.8 & 218 & 1.84 & 154 & 1.30 & 1.9 \\
NGC 2808 & 282.19 & -11.25 & 10.2 & 543 & 1.69 & 347 & 1.08 & 0.8 \\
NGC 6553 &  5.25 & -3.03 & 6.8 & 198 & 1.66 & 198 & 1.66 & 1.4 \\
NGC 6341 & 68.34 & 34.86 & 8.4 & 345 & 1.63 & 190 & 0.90 & 0.2 \\
NGC 6293 & 357.62 &  7.83 & 9.2 & 332 & 1.62 & 17 & 0.08 & 0.5 \\
NGC 5946 & 327.58 &  4.19 & 10.6 & 272 & 1.46 & 126 & 0.68 & 0.3 \\
Ter 5 &  3.84 &  1.69 & 5.5 & 151 & 1.45 & 73 & 0.70 & 1.1 \\
NGC 6218 & 15.71 & 26.31 & 4.7 & 120 & 1.44 & 100 & 1.20 & 2.4 \\
NGC 6642 &  9.81 & -6.44 & 8.1 & 198 & 1.40 & 40 & 0.28 & 0.1 \\
NGC 7078 & 65.01 & -27.31 & 10.2 & 445 & 1.39 & 472 & 1.47 & 4.0 \\
NGC 362 & 301.53 & -46.25 & 9.2 & 244 & 1.37 & 132 & 0.74 & 0.3 \\
NGC 6256 & 347.79 &  3.31 & 6.4 & 200 & 1.34 & 127 & 0.85 & 2.3 \\
NGC 6284 & 358.35 &  9.94 & 15.1 & 366 & 1.32 & 173 & 0.63 & 0.8 \\
NGC 6539 & 20.80 &  6.78 & 7.8 & 233 & 1.29 & 186 & 1.03 & 1.4 \\
NGC 6397 & 338.17 & -11.96 & 2.4 & 58 & 1.26 & 70 & 1.53 & 2.1 \\
NGC 6325 &  0.97 &  8.00 & 7.8 & 171 & 1.23 & 25 & 0.18 & 0.4 \\
NGC 7089 & 53.37 & -35.77 & 10.5 & 301 & 1.20 & 154 & 0.61 & 0.6 \\
NGC 6541 & 349.29 & -11.19 & 8.0 & 261 & 1.20 & 222 & 1.02 & 1.6 \\
Ter 2 & 356.32 &  2.30 & 7.5 & 348 & 1.09 & 35 & 0.11 & 0.3 \\
Ter 4 & 356.02 &  1.31 & 6.7 & 128 & 1.02 & 5 & 0.04 & 0.6 \\
    \end{tabular}
\end{table*}
\begin{table*}
    \centering
    \caption{Continuation of Table~\ref{tab:app1}}
    \begin{tabular}{l|rrrrrrrr}
    \hline
GC & $l$ & $b$ &  $D_\odot$  & $\Delta x$ (isotropic) & $\Delta \phi$ (isotropic) & $\Delta x$ & $\Delta \phi$ & Pericentre \\
 & deg & deg & kpc & pc & deg & pc & deg & kpc \\
 \hline

NGC 5286 & 311.61 & 10.57 & 11.4 & 224 & 0.96 & 125 & 0.54 & 0.9 \\
NGC 4372 & 300.99 & -9.88 & 5.8 & 236 & 0.95 & 239 & 0.96 & 2.9 \\
NGC 6356 &  6.72 & 10.22 & 15.1 & 265 & 0.87 & 313 & 1.03 & 2.4 \\
NGC 6779 & 62.66 &  8.34 & 9.7 & 162 & 0.84 & 106 & 0.55 & 0.5 \\
NGC 6624 &  2.79 & -7.91 & 7.2 & 118 & 0.77 & 2 & 0.01 & 0.4 \\
Ter 10 &  4.42 & -1.87 & 5.8 & 79 & 0.76 & 9 & 0.09 & 0.8 \\
NGC 6540 &  3.29 & -3.31 & 5.2 & 68 & 0.75 & 47 & 0.52 & 1.8 \\
NGC 6254 & 15.14 & 23.08 & 5.0 & 189 & 0.75 & 178 & 0.70 & 1.9 \\
NGC 5897 & 342.95 & 30.29 & 12.6 & 170 & 0.74 & 173 & 0.75 & 3.1 \\
NGC 6681 &  2.85 & -12.51 & 9.3 & 204 & 0.69 & 61 & 0.21 & 0.0 \\
NGC 5904 &  3.86 & 46.80 & 7.6 & 91 & 0.67 & 124 & 0.91 & 5.4 \\
NGC 6139 & 342.37 &  6.94 & 9.8 & 117 & 0.66 & 52 & 0.29 & 1.1 \\
NGC 6352 & 341.42 & -7.17 & 5.9 & 119 & 0.58 & 90 & 0.43 & 3.1 \\
FSR 1735 & 339.19 & -1.85 & 9.8 & 128 & 0.57 & 49 & 0.22 & 0.6 \\
IC 1276 & 21.83 &  5.67 & 5.4 & 141 & 0.56 & 126 & 0.50 & 3.5 \\
UKS 1 &  5.12 &  0.76 & 7.8 & 104 & 0.56 & 37 & 0.20 & 0.6 \\
NGC 6652 &  1.53 & -11.38 & 10.0 & 108 & 0.55 & 15 & 0.08 & 0.3 \\
NGC 6838 & 56.75 & -4.56 & 4.0 & 89 & 0.50 & 79 & 0.45 & 5.2 \\
NGC 6528 &  1.14 & -4.17 & 7.5 & 302 & 0.49 & 60 & 0.10 & 0.5 \\
NGC 6366 & 18.41 & 16.04 & 3.7 & 31 & 0.48 & 38 & 0.59 & 1.9 \\
NGC 6760 & 36.11 & -3.92 & 8.0 & 178 & 0.47 & 214 & 0.56 & 2.1 \\
NGC 5024 & 332.96 & 79.76 & 17.9 & 158 & 0.46 & 163 & 0.48 & 14.6 \\
NGC 6723 &  0.07 & -17.30 & 8.3 & 73 & 0.46 & 99 & 0.62 & 1.1 \\
NGC 6558 &  0.20 & -6.02 & 7.2 & 113 & 0.43 & 24 & 0.09 & 0.4 \\
Pal 8 & 14.11 & -6.80 & 12.8 & 95 & 0.41 & 92 & 0.40 & 2.2 \\
Djor 2 &  2.76 & -2.51 & 6.3 & 54 & 0.41 & 1 & 0.00 & 1.0 \\
NGC 6362 & 325.56 & -17.57 & 7.4 & 136 & 0.41 & 123 & 0.37 & 2.2 \\
NGC 5927 & 326.60 &  4.86 & 8.2 & 463 & 0.40 & 392 & 0.34 & 4.2 \\
NGC 6171 &  3.37 & 23.01 & 6.0 & 42 & 0.39 & 51 & 0.48 & 1.1 \\
NGC 1851 & 244.51 & -35.04 & 11.3 & 194 & 0.39 & 145 & 0.29 & 0.1 \\
2MASS-GC02 0 &  9.78 & -0.62 & 7.1 & 96 & 0.38 & 26 & 0.10 & 0.5 \\
NGC 6287 &  0.13 & 11.02 & 9.4 & 134 & 0.37 & 45 & 0.13 & 0.3 \\
NGC 5824 & 332.56 & 22.07 & 31.8 & 238 & 0.35 & 271 & 0.40 & 10.0 \\
NGC 6333 &  5.54 & 10.71 & 8.4 & 194 & 0.35 & 55 & 0.10 & 0.8 \\
FSR 1716 & 329.78 & -1.59 & 7.5 & 65 & 0.35 & 57 & 0.31 & 2.3 \\
Pal 10 & 52.44 &  2.73 & 5.9 & 87 & 0.35 & 78 & 0.31 & 4.2 \\
NGC 1904 & 227.23 & -29.35 & 13.3 & 114 & 0.35 & 89 & 0.27 & 0.2 \\
Ter 9 &  3.60 & -1.99 & 7.1 & 63 & 0.34 & 6 & 0.03 & 0.2 \\
BH 261 &  3.36 & -5.27 & 6.5 & 39 & 0.34 & 26 & 0.23 & 1.5 \\
Lynga 7 & 328.77 & -2.80 & 8.0 & 61 & 0.33 & 73 & 0.39 & 1.7 \\
Ter 3 & 345.08 &  9.19 & 8.1 & 77 & 0.32 & 56 & 0.24 & 2.0 \\
NGC 6637 &  1.72 & -10.27 & 8.8 & 56 & 0.32 & 30 & 0.17 & 0.7 \\
NGC 3201 & 277.23 &  8.64 & 4.5 & 37 & 0.32 & 46 & 0.39 & 7.8 \\
Pal 11 & 31.80 & -15.58 & 14.3 & 87 & 0.32 & 81 & 0.29 & 5.6 \\
NGC 6144 & 351.93 & 15.70 & 8.9 & 75 & 0.27 & 58 & 0.21 & 1.9 \\
NGC 288 & 151.28 & -89.38 & 10.0 & 51 & 0.25 & 67 & 0.33 & 3.7 \\
Ton 2 & 350.80 & -3.42 & 6.4 & 34 & 0.21 & 27 & 0.16 & 1.9 \\
NGC 1261 & 270.54 & -52.12 & 15.5 & 88 & 0.21 & 101 & 0.25 & 5.7 \\
NGC 7099 & 27.18 & -46.84 & 8.0 & 46 & 0.19 & 75 & 0.31 & 1.6 \\
NGC 6342 &  4.90 &  9.72 & 8.4 & 37 & 0.11 & 31 & 0.09 & 0.9 \\
    \end{tabular}
\end{table*}
\begin{table*}
    \centering
    \caption{Continuation of Table~\ref{tab:app1}}
    \begin{tabular}{l|rrrrrrrr}
    \hline
GC & $l$ & $b$ &  $D_\odot$  & $\Delta x$ (isotropic) & $\Delta \phi$ (isotropic) & $\Delta x$ & $\Delta \phi$ & Pericentre \\
 & deg & deg & kpc & pc & deg & pc & deg & kpc \\
 \hline
NGC 6496 & 348.03 & -10.01 & 11.3 & 26 & 0.11 & 31 & 0.13 & 3.9 \\
NGC 6235 & 358.92 & 13.52 & 13.5 & 27 & 0.11 & 38 & 0.15 & 4.9 \\
NGC 6864 & 20.30 & -25.75 & 21.6 & 82 & 0.10 & 139 & 0.17 & 3.4 \\
NGC 5053 & 335.70 & 78.95 & 17.2 & 30 & 0.10 & 31 & 0.10 & 12.7 \\
NGC 6717 & 12.88 & -10.90 & 7.1 & 21 & 0.09 & 14 & 0.06 & 1.3 \\
NGC 6316 & 357.18 &  5.76 & 11.6 & 21 & 0.09 & 132 & 0.53 & 1.2 \\
NGC 6749 & 36.19 & -2.21 & 7.8 & 26 & 0.08 & 45 & 0.14 & 1.7 \\
NGC 6981 & 35.16 & -32.68 & 17.0 & 47 & 0.08 & 51 & 0.09 & 5.5 \\
NGC 2298 & 245.63 & -16.01 & 10.8 & 19 & 0.08 & 10 & 0.04 & 1.3 \\
NGC 6584 & 342.14 & -16.41 & 13.2 & 20 & 0.07 & 60 & 0.21 & 2.2 \\
NGC 6229 & 73.64 & 40.31 & 30.6 & 89 & 0.06 & 70 & 0.05 & 0.7 \\
ESO 452-SC11 & 351.91 & 12.10 & 6.5 & 7 & 0.06 & 1 & 0.01 & 0.5 \\
NGC 6809 &  8.79 & -23.27 & 5.3 & 38 & 0.06 & 43 & 0.06 & 1.1 \\
NGC 5634 & 342.21 & 49.26 & 27.2 & 39 & 0.05 & 58 & 0.08 & 5.1 \\
NGC 6101 & 317.75 & -15.82 & 16.1 & 17 & 0.05 & 21 & 0.06 & 11.7 \\
NGC 6535 & 27.18 & 10.44 & 6.5 & 6 & 0.04 & 1 & 0.00 & 1.0 \\
NGC 4590 & 299.63 & 36.05 & 10.1 & 9 & 0.03 & 13 & 0.05 & 9.1 \\
Ter 12 &  8.36 & -2.10 & 3.4 & 2 & 0.02 & 2 & 0.02 & 3.1 \\
NGC 5466 & 42.15 & 73.59 & 16.9 & 8 & 0.02 & 10 & 0.03 & 12.9 \\
NGC 5694 & 331.06 & 30.36 & 37.3 & 40 & 0.02 & 49 & 0.03 & 13.9 \\
Pal 5 &  0.85 & 45.86 & 23.2 & 8 & 0.02 & 8 & 0.02 & 16.3 \\
ESO 280-SC06 & 346.90 & -12.57 & 22.9 & 9 & 0.02 & 3 & 0.01 & 1.7 \\
Pal 2 & 170.53 & -9.07 & 27.2 & 34 & 0.02 & 25 & 0.01 & 2.2 \\
NGC 6934 & 52.10 & -18.89 & 15.4 & 5 & 0.02 & 2 & 0.01 & 4.7 \\
NGC 7492 & 53.39 & -63.48 & 26.6 & 11 & 0.01 & 12 & 0.02 & 10.1 \\
NGC 4147 & 252.85 & 77.19 & 18.2 & 15 & 0.01 & 11 & 0.01 & 0.3 \\
IC 1257 & 16.53 & 15.14 & 25.0 & 9 & 0.01 & 1 & 0.00 & 2.3 \\
E 3 & 292.27 & -19.02 & 8.1 & 2 & 0.01 & 2 & 0.01 & 8.7 \\
IC 4499 & 307.35 & -20.47 & 18.2 & 3 & 0.01 & 11 & 0.03 & 5.0 \\
NGC 2419 & 180.37 & 25.24 & 83.2 & 7 & 0.00 & 21 & 0.01 & 10.3 \\
Pal 1 & 130.06 & 19.03 & 11.0 & 1 & 0.00 & 1 & 0.00 & 15.4 \\
Ter 7 &  3.39 & -20.07 & 22.8 & 1 & 0.00 & 0 & 0.00 & 3.8 \\
Djor 1 & 356.68 & -2.48 & 13.7 & 1 & 0.00 & 1 & 0.00 & 5.6 \\
Rup 106 & 300.89 & 11.67 & 21.2 & 1 & 0.00 & 0 & 0.00 & 4.6 \\
Ter 8 &  5.76 & -24.56 & 26.7 & 0 & 0.00 & 2 & 0.00 & 6.5 \\
Pyxis 0 & 261.32 &  7.00 & 39.4 & 0 & 0.00 & 0 & 0.00 & 5.0 \\
Pal 15 & 18.85 & 24.34 & 45.1 & 2 & 0.00 & 1 & 0.00 & 3.8 \\
Pal 12 & 30.51 & -47.68 & 19.0 & 0 & 0.00 & 0 & 0.00 & 14.5 \\
NGC 7006 & 63.77 & -19.41 & 42.8 & 1 & 0.00 & 5 & 0.00 & 7.4 \\
Pal 3 & 240.14 & 41.87 & 92.5 & 0 & 0.00 & 0 & 0.00 & 58.7 \\
Whiting 1 & 161.62 & -60.64 & 31.3 & 0 & 0.00 & 0 & 0.00 & 29.6 \\
Arp 2 &  8.54 & -20.79 & 28.6 & 0 & 0.00 & 1 & 0.00 & 6.2 \\
Eridanus 0 & 218.11 & -41.33 & 90.1 & 0 & 0.00 & 0 & 0.00 & 56.1 \\
AM 1 & 258.36 & -48.47 & 123.3 & 0 & 0.00 & 0 & 0.00 & 102.3 \\
Pal 13 & 87.10 & -42.70 & 24.8 & 0 & 0.00 & 0 & 0.00 & 10.9 \\
Crater 0 & 274.81 & 47.85 & 145.0 & 0 & 0.00 & 0 & 0.00 & 90.4 \\
NGC 6426 & 28.09 & 16.23 & 35.3 & 0 & 0.00 & 0 & 0.00 & 28.8 \\
AM 4 & 320.28 & 33.51 & 32.2 & 0 & 0.00 & 0 & 0.00 & 12.3 \\
Pal 14 & 28.75 & 42.19 & 71.0 & 0 & 0.00 & 0 & 0.00 & 10.3 \\
Pal 4 & 202.31 & 71.80 & 103.0 & 0 & 0.00 & 0 & 0.00 & 10.2 \\
\hline
    \end{tabular}
\end{table*}

\bsp	
\label{lastpage}
\end{document}